\documentclass[aps,pre,reprint]{revtex4-1}

\usepackage[utf8]{inputenc} 
\usepackage[T1]{fontenc}    %
\usepackage[english]{babel} 

\usepackage{amssymb,amsmath,mathtools} 
\usepackage{graphicx} 

\renewcommand{\Im}{\operatorname{Im}} 

\newcommand{\nn}{\nonumber} 

\newcommand{\C}{\mathbb{C}} %
\newcommand{\E}{\mathbb{E}} %

\newcommand{\cI}{\mathcal I}

\newcommand{\cN}{\mathcal N}

\newcommand{\cP}{\mathcal P}
\newcommand{\cU}{\mathcal U}

\newcommand{\eins}{\leavevmode\hbox{\small1\kern-3.8pt\normalsize1}}

\newcommand{\mX}{\mathbb X}
\newcommand{\mY}{\mathbb Y}

\DeclareMathOperator{\gO}{O}
\DeclareMathOperator{\gU}{U}
\DeclareMathOperator{\gSp}{USp}

\let\olddet\det
\renewcommand{\det}{\olddet\nolimits} 

\renewcommand{\phi}{\varphi} 
\renewcommand{\epsilon}{\varepsilon} 

\DeclareMathOperator{\tr}{Tr} 
\DeclareMathOperator*{\diag}{diag} 

\DeclarePairedDelimiter{\abs}{\lvert}{\rvert} 

\newcommand{\jpdf}{\text{jpdf}}

\newcommand{\MeijerG}[8][\Big]{G^{{ #2 },\,{ #3 }}_{{ #4 },\,{ #5 }} #1( \begin{smallmatrix} #6 \\ #7 \end{smallmatrix}\, #1\vert\, #8 #1)}

\newcommand{\hypergeometric}[6][\Big]{\,{}_{#2} F_{#3} #1( \begin{smallmatrix} #4 \\ #5 \end{smallmatrix}\, #1\vert\, #6 #1)}

\usepackage{hyperref}   

\begin{document}

\title{Products of Rectangular Random Matrices:\\ Singular Values and Progressive Scattering}
\author{Gernot Akemann}\email{akemann@physik.uni-bielefeld.de}\author{Jesper R. Ipsen}\email{jipsen@math.uni-bielefeld.de}\author{Mario Kieburg}\email{mkieburg@physik.uni-bielefeld.de}
\affiliation{Department of Physics, Bielefeld University, Postfach 100131, D-33501 Bielefeld, Germany}
\date{\today}

\begin{abstract}
We discuss the product of $M$ rectangular random matrices with independent Gaussian entries, which have several applications including wireless telecommunication and econophysics. For complex matrices an explicit expression for the joint probability density function is obtained using the Harish-Chandra--Itzykson--Zuber integration formula. Explicit expressions for all correlation functions and moments for finite matrix sizes are obtained using a two-matrix model and the method of bi-orthogonal polynomials. This generalises the classical result for the so-called Wishart--Laguerre Gaussian unitary ensemble (or chiral unitary ensemble) at $M=1$, and previous results for the product of square matrices. The correlation functions are given by a determinantal point process, where the kernel can be expressed in terms of Meijer $G$-functions. We compare the results with numerical simulations and known results for the macroscopic level density in the limit of large matrices. The location of the endpoints of support for the latter are analysed in detail for general $M$. Finally, we consider the so-called ergodic mutual information, which gives an upper bound for the spectral efficiency of a MIMO communication
channel with multi-fold scattering.
\end{abstract}

\maketitle


\section{Introduction}
\label{sec:intro}

Random Matrix Theory has existed for more than half a century, and its success is undeniable. A vast number of applications is known within the mathematical and physical sciences, and beyond; we refer to~\cite{Oxford:2011} for a recent overview. A direction within Random Matrix Theory, which has recently caught renewed attention is the study of products of random matrices. Among others, products of matrices have been applied to disordered and chaotic systems~\cite{CPV:1993}, matrix-valued diffusions~\cite{JLJN:2002,GJJN:2003}, quantum chromodynamics at finite chemical potential~\cite{Osborn:2004,Akemann:2007}, Yang--Mills theory~\cite{BLNT:2005,NN:2007,BN:2008}, finance~\cite{Bouchaud} and wireless telecommunication~\cite{TV:2004}. In this paper, our attention will be directed towards the latter.

When considering products of matrices we are faced with the fact that the product often possesses less symmetries than the individual matrices. For example a product of symmetric matrices  will not be symmetric in general. For simplicity, we will look at matrices with a minimum of symmetry. Our discussion will concern products of matrices drawn from the Wishart ensemble. Thus the matrices have independently, identically distributed Gaussian entries. Also other proposals exist, e.g. by multiplying matrices that are chosen from a 
set of {\it fixed} matrices with a given probability. This problem 
has applications in percolation as was pointed out in \cite{Mehtabook}. 
However it considerably differs from our approach, notably due to the lack 
of invariance.

The statistical properties of the complex eigenvalues and real singular values of a product of matrices from the Wishart ensemble have been discussed in several papers (in the former case they are usually called Ginibre matrices). Macroscopic properties for eigenvalues of complex ($\beta=2$) matrices have been discussed in the limit of large matrices using diagrammatic methods~\cite{GJJN:2003,BJW:2010,BJLNS:2010}, while proofs are given in~\cite{GT:2010,RS:2011}. The macroscopic behaviour of the singular values and their moments have also been discussed in the literature using probabilistic methods~\cite{Muller:2002,BBCC:2007,BG:2008} as well as diagrammatic methods~\cite{BJLNS:2010}.

Recently, the discussion of products of matrices from Wishart ensembles has been extended to matrices of finite size~\cite{AB:2012,AS:2012,Ipsen:2013,AKW:2013}, but this discussion has so far been limited to the case of square matrices. We want to extend this discussion to include products of rectangular matrices. In particular, we consider the product matrix
\begin{equation}
\mY_M=\mX_M \mX_{M-1} \cdots \mX_1,
\label{intro:product}
\end{equation}
where $\mX_m$ are $N_m\times N_{m-1}$ real ($\beta=1$), complex ($\beta=2$) or quaternion ($\beta=4$) matrices from the Wishart ensemble. This paper is concerned with  the singular values of such matrices, and the spectral correlation functions of $\mY_M\mY_M^\dag$. A discussion of the complex eigenvalues is postponed to a future publication~\cite{IK:2013}.

Matrix products like $\mathbb{Y}_M$ have direct applications in finance~\cite{Bouchaud} , wireless telecommunication~\cite{Muller:2002} and quantum entanglement~\cite{BNZ:2010,ZPNC:2011}.
The importance of the generalisation from square to rectangular matrices is evident from its applications to e.g. wireless telecommunication. Let us consider a MIMO (Multiple--Input Multiple--Output) communication channel from a single source to a single destination via $M-1$ clusters of scatterers. The source and destination are assumed to be equipped with $N_0$ transmitting and $N_M$ receiving antennas, respectively. Each cluster of scatterers is assumed to have $N_m$ ($1\leq m\leq M-1$) scattering objects. Such a communication link is canonically described by a channel matrix identical to the complex version of the product matrix~\eqref{intro:product}. Here the Gaussian nature of the matrix entries models a Rayleigh fading environment. This model was proposed in~\cite{Muller:2002}, while the single channel model ($M=1$) goes back to~\cite{Foschini:1996,FG:1998,Telatar:1999}. There is no reason to assume that the number of scattering object at each cluster in such a communication channel should be
identical, which illustrates the importance
of the generalisation to rectangular matrices.

This paper will be organised as follows: In section~\ref{sec:jpdf} we will find the joint probability density function for the singular values of the product matrix~\eqref{intro:product} in the complex case. Starting with general $\beta=1,2,4$ it turns out that the restriction to complex ($\beta=2$) matrices is necessary, since our method relies on the Harish-Chandra--Itzykson--Zuber integration formula for the unitary group~\cite{HC:1957,IZ:1980}. An explicit expression for all $k$-point correlation functions for the singular values will be derived in section~\ref{sec:2mm} using a two-matrix model and the method of bi-orthogonal polynomials. The spectral density and its moments will be discussed further in section~\ref{sec:moment}, while we return to the above mentioned communication channel in section~\ref{sec:mimo}. Section~\ref{sec:conclusion} is devoted to conclusions and outlook. Some properties and identities for the special functions we encounter are collected in appendix~\ref{sec:meijer}.

\section{Joint Probability Distribution of Singular Values}
\label{sec:jpdf}

As mentioned in the introduction we are interested in the statistical properties of the singular values of the product matrix~\eqref{intro:product}, which is governed by the following partition function,
\begin{equation}
Z^M_\beta=\prod_{m=1}^M\int\abs{D\mX_m}\exp[-\tr \mX_m\mX_m^\dagger].
\end{equation}
Here $D\mX_m$ denotes the Euclidean volume, i.e. the exterior product of all independent one-forms, while $\abs{D\mX_m}$ is the corresponding unoriented volume element.

Let us assume that the smallest dimension is $N_0=N_{\min}$. We stress that the properties of the non-zero singular values of $\mY_M$ are completely independent of this choice, see \cite{IK:2013}. Thus, the product matrix, $\mY_M=\mX_M\cdots\mX_1$, has maximally rank $N_0$. It follows that the product matrix can be parameterised as~\cite{IK:2013}
\begin{equation}
\mY_M=\cU_M
\begin{pmatrix}
Y_M \\
0 
\end{pmatrix}
,
\label{jpdf:parameterization}
\end{equation}
where $Y_M$ is a square $N_0\times N_0$ matrix with real, complex or quaternion entries, while $\cU_M$ is an orthogonal, a unitary or a unitary symplectic matrix for $\beta=1,2,4$, respectively. From equation~\eqref{jpdf:parameterization} it is immediate that the non-zero singular values of the rectangular matrix $\mY_M$ are identical to the singular values of the square matrix $Y_M$.  The ultimate goal is to derive the joint probability density function for these singular values. In~\cite{IK:2013} the invariance of the matrix measure for $Y_M$  under permutations of the matrix dimensions, $N_m$, was shown This invariance carries over to the  joint probability density function of the singular values as we will see.

The parametrisation~\eqref{jpdf:parameterization} follows directly from a parametrisation of each individual matrix,
\begin{equation}
\mX_m=\cU_m
\begin{pmatrix}
X_m & A_m\\
0   & B_m
\end{pmatrix}
\cU_{m-1}^{-1},
\end{equation}
where $\cU_0=\eins_{N_0}$. The matrices $X_m$, $A_m$ and $B_m$ have the dimensions $N_{0}\times N_{0}$, $N_{0}\times (N_{m-1}-N_{0})$ and $(N_m-N_{0})\times (N_{m-1}-N_{0})$, respectively. The entries of these matrices are real for $\beta=1$, complex for $\beta=2$ and quaternion for $\beta=4$. Accordingly, we have
\begin{equation}
\cU_m\in
\begin{cases}
 \gO(N_m)/[\gO(N_{0})\times \gO(N_m-N_{0})],\\
 \gU(N_m)/[\gU(N_{0})\times \gU(N_m-N_{0})],\\
 \gSp(2N_m)/[\gSp(2N_{0})\times \gSp(2(N_m-N_{0}))],
\end{cases}
\end{equation}
for $\beta=1,2,4$, respectively. The non-zero singular values of the rectangular product matrix~\eqref{intro:product} are identical to the singular values of the square product matrix $Y_M=X_MX_{M-1}\cdots X_1$ with $Y_M$ and $X_m$, $m=1,\ldots,M$, defined above. For this reason, we can safely replace the random matrix model containing rectangular matrices with a random matrix model containing square matrices, only.  
In terms of the new variables we get for the partition function, in analogy to \cite{FBKSZ:2012} for $M=1$, 
\begin{equation}
Z_\beta^M\propto\prod_{m=1}^M\int \abs{DX_m} \det^{\beta\nu_m/2}(X_mX_m^\dagger)\exp[-\tr X_mX_m^\dagger],
\label{jpdf:ZX}
\end{equation}
where $\nu_m\equiv N_m-N_{0}\geq 0$. A more general version of this result will be derived in~\cite{IK:2013}. In the partition function~\eqref{jpdf:ZX} and in most of this section we neglect an overall normalisation constant, which is irrelevant for the computations. We reintroduce the normalisation in equation~\eqref{jpdf:Zs2} and give the explicit value in equation~\eqref{jpdf:norm}.

The Gaussian weight times a determinantal prefactor is sometimes referred to as the induced weight. For $M=1$ its complex eigenvalues have been studied in~\cite{FBKSZ:2012}.

In order to derive the joint probability density function for the singular values of the product matrix $Y_M$ and thereby of equation~\eqref{intro:product}, we follow the idea in~\cite{AKW:2013}, and reformulate the partition function~\eqref{jpdf:ZX} in terms of the product matrices $Y_m=X_mY_{m-1}=X_mX_{m-1}\cdots X_1$, for $m=1,\ldots,M$. In the following we assume that the product matrices, $Y_m$, are invertible (note that this restriction only removes a set of measure zero).
We then know that~\cite{AKW:2013}
\begin{equation}
\prod_{m=1}^M\abs{DX_m}=\abs{DY_1}\prod_{m=2}^M \abs{DY_m}\det^{-\beta N_{0}/2}(Y_{m-1}Y_{m-1}^\dagger).
\end{equation}
Changing variables from $X_m$ to $Y_m$ in the partition function equation~\eqref{jpdf:ZX} results in
\begin{align}
Z_\beta^M \propto \Big[\prod_{m=1}^{M}\int &\!\abs{DY_m}\Big] \det^{\beta\nu_M/2}(Y_MY_M^\dagger) 
\exp\left[-\tr Y_{1}Y_{1}^\dagger\right] \nn\\
&\times\left[\prod_{i=2}^{M}\det^{\beta(\nu_{i-1}-\nu_{i}-N_{0})/2}(Y_{i-1}Y_{i-1}^\dagger)\right. \nn\\
&\times\left.\exp\left[-\tr Y_{i}Y_{i}^\dagger(Y_{i-1}Y_{i-1}^\dagger)^{-1}\right]\right].
\label{jpdf:ZY}
\end{align}
With this expression for the partition function we can express everything in terms of the singular values and a family of unitary matrices. We employ for each matrix $Y_i$ a singular value decomposition~\cite{AKW:2013} to write the product matrices as
\begin{equation}
Y_i=U_i\Sigma_iV_i^{-1},
\end{equation}
where $\Sigma_i=\diag\{\sigma_{1}^i,\sigma_{2}^i,\ldots,\sigma_{N_{0}}^i\}$ are positive definite diagonal matrices; the diagonal elements are the singular values of $Y_i$ (for $\beta=4$ the singular values show Kramer's degeneracy). The unitary matrices, $U_i$ and $V_i$, belong to
\begin{equation}
U_i\in
\begin{cases}
\gO(N_{0}),\\
\gU(N_{0}),\\
\gSp(2N_{0}),
\end{cases}\quad
V_i\in
\begin{cases}
\gO(N_{0}),\\
\gU(N_{0})/\gU(1)^{N_{0}},\\
\gSp(2N_{0})/\gU(1)^{N_{0}},
\end{cases}
\end{equation}
for $\beta=1,2,4$, respectively. It is well-known that this change of variables yields the new measure
\begin{equation}
\abs{DY_i}=\abs{DU_i}\abs{DV_i} \prod_{k=1}^{N_0} d\sigma_k^i(\sigma_k^i)^{\beta-1} \abs{\Delta_{N_{0}}((\sigma^i)^2)}^\beta ,
\end{equation}
where $\abs{DU_i}$ and $\abs{DV_i}$ are the Haar measures for their corresponding groups and
\begin{equation}
\Delta_N(x)=\prod_{1\leq a<b\leq N}(x_a-x_b)=\olddet_{1\leq a,b\leq N}[x_a^{N-b}]
\end{equation}
denotes the Vandermonde determinant.
Inserting this parametrisation into the partition function~\eqref{jpdf:ZY} and performing the shift $U_{\ell-1}^{-1}U_\ell\to U_\ell$ for $\ell=2,\ldots,M$, we obtain
\begin{align}
Z_\beta^M &\propto\left[ \prod_{k=1}^{N_0}\bigg[\prod_{m=1}^M\int_0^\infty d\sigma_k^m\bigg]
(\sigma_k^M)^{\beta(\nu_M+1)-1}e^{-(\sigma_k^1)^2} \right.\nn\\
&\times\left.\prod_{i=2}^M (\sigma_k^{i-1})^{\beta(\nu_{i-1}-\nu_i-N_{0}+1)-1}\right]\prod_{j=1}^M\abs{\Delta_{N_{0}}((\sigma^j)^2)}^\beta  \nn\\
&\times\prod_{\ell=2}^M  \int \abs{DU_\ell}\abs{DV_\ell}\exp\Big[-\tr U_\ell\Sigma_\ell^2 U_\ell^{-1} \Sigma_{\ell-1}^{-2}\Big].
\label{jpdf:ZUV}
\end{align}
The integrations over $V_\ell$ are trivial and only contribute to the normalisation constant; the integration over $U_\ell$ is however more complicated. For $\beta=2$, the integrals over $U_\ell$ are Harish-Chandra--Itzykson--Zuber integrals~\cite{HC:1957,IZ:1980}, while the integrals for $\beta=1$ and $\beta=4$ are still unknown in closed form. For this reason, we will restrict ourselves to the complex case ($\beta=2$), where we can carry out all integrals explicitly, and obtain an analytical expression for the joint probability density function. Recall that the complex ($\beta=2$) product matrix is exactly the channel matrix used in wireless telecommunication to model MIMO channels with multiple scattering.

With the restriction to the $\beta=2$ case, $U_\ell$ should be integrated over the unitary group, which yields~\cite{HC:1957,IZ:1980}
\begin{multline}
\int_{\gU(N_{0})}\abs{DU_\ell}\exp\Big[-\tr U_\ell\Sigma_\ell^2 U_\ell^{-1} \Sigma_{\ell-1}^{-2}\Big]\propto\\
\frac{\prod_{k=1}^{N_0}(\sigma_k^{\ell-1})^{2(N_{0}-1)}}{\Delta_{N_{0}}((\sigma^\ell)^2)\Delta_{N_{0}}((\sigma^{\ell-1})^2)}
\olddet_{1\leq a,b\leq N_{0}}\left[e^{-(\sigma^\ell_{a})^2/(\sigma^{\ell-1}_b)^2}\right],
\end{multline}
for $\ell=2,\ldots,M$.
Inserting this into the partition function~\eqref{jpdf:ZUV} with $\beta=2$ gives an expression for the partition function solely in terms of the singular values of the product matrices $Y_i$,
\begin{align}
Z^M  &\equiv Z_{\beta=2}^M \propto 
\left[\prod_{k=1}^{N_0}\int_0^\infty d\sigma_k^M(\sigma_k^M)^{2\nu_M+1}\right]\Delta_{N_{0}}((\sigma^M)^2) \nn\\
&\times\left[\prod_{i=1}^{M-1}\bigg[\prod_{\ell=1}^{N_0}\int_0^\infty d\sigma_\ell^i (\sigma_\ell^{i})^{2(\nu_{i}-\nu_{i+1})-1} \bigg]\right. \nn\\
&\times \left.\olddet_{1\leq a,b\leq N_{0}}\!\left[e^{-(\sigma^{i+1}_{a})^2/(\sigma^{i}_{b})^2}\right]\right]  \left[\prod_{k=1}^{N_0}e^{-(\sigma_k^1)^2}\right]\nn\\
&\times\Delta_{N_{0}}((\sigma^1)^2).
\label{jpdf:Zs}
\end{align}
For notational simplicity we will change variables from the singular values to $s_{a}^i=(\sigma_{a}^i)^2$, i.e. the singular values (and eigenvalues) of the Wishart matrices $Y_iY_i^\dagger$ (the singular values of $Y_MY_M^\dagger$ will simply be denoted by $s_a=s_a^M$). Furthermore, due to symmetrisation we can replace the determinants of the exponentials by their diagonals, which will only change the partition function by a factor $(N_{0}!)^{M-1}$. Exploiting this, the partition function becomes
\begin{align}
Z^M&=C_M^{-1}\left[ \prod_{b=1}^{N_{0}} \int_0^\infty ds_{b}\, (s_{b})^{\nu_M}\right]\, \Delta_{N_{0}}(s) \nn\\
&\times\left[ \prod_{a=1}^{N_{0}}\bigg[\prod_{i=1}^{M-1} \int_0^\infty \frac{ds_{a}^i}{s_{a}^i}\, (s_{a}^i)^{\nu_i-\nu_{i+1}}\,e^{-s_a^{i+1}/s_{a}^i}\bigg]
e^{-s_{a}^1}\right]\nn\\ 
&\times\Delta_{N_{0}}(s^1),
\label{jpdf:Zs2}
\end{align}
where $C_M$ is a normalisation constant.

The integrations over $s_a^1,\ldots,s_a^{M-1}$ have a similar structure. Hence, we can perform all these integrals in a similar fashion. We write the first exponential containing $s_a^1$ as a Meijer $G$-function using equation~\eqref{meijer:meijer-func}, i.e.
\begin{equation}
\Delta_{N_{0}}(s^1)\prod_{a=1}^{N_0}e^{-s_{a}^1}=\olddet_{1\leq a,b\leq N_0}\Big[\MeijerG{1}{0}{0}{1}{-}{b-1}{s_a^1}\Big].
\end{equation}
After a change of variables all the integrals can be performed inductively using the identities~\eqref{meijer:meijer-shift} and~\eqref{meijer:meijer-induc}. These integrations finally give the joint probability density function, $\cP_\jpdf$, for the singular values $s_{1},\ldots,s_{N_{0}}$ of the Wishart matrix $Y_MY_M^\dagger$,
\begin{multline}
\cP_{\jpdf}^M(s_1,\ldots,s_{N_{0}})=
C_M^{-1}\Delta_{N_{0}}(s)\\
\times\olddet_{1\leq a,b\leq N_{0}}\left[\MeijerG{M}{0}{0}{M}{-}{\nu_M,\,\nu_{M-1},\,\ldots\,,\,\nu_2,\,\nu_1+b-1}{s_a}\right].
\label{jpdf:jpdf}
\end{multline}
The partition function is thus given by
\begin{equation}
Z^M=\prod_{a=1}^{N_{0}}\int_0^\infty ds_a\,\cP_\jpdf^M(s_1,\ldots,s_{N_{0}}).
\label{jpdf:Zjpdf}
\end{equation}
This generalises the joint probability density function for the product of square matrices from the Wishart ensemble given in~\cite{AKW:2013} to the case of rectangular matrices. In principle all $k$-point correlation functions for the singular values, $R_k^M(s_1,\ldots,s_k)$, can be calculated from the joint probability density function~\eqref{jpdf:jpdf} as
\begin{multline}
R_k^M(s_1,\ldots,s_k)=\\
\frac{N_0!}{(N_0-k)!}\prod_{a=k+1}^{N_0}\int_0^\infty ds_a\, \cP_{\jpdf}^M(s_1,\ldots,s_{N_{0}}).
\label{jpdf:R}
\end{multline}
Due to the Meijer $G$-function inside the determinant~\eqref{jpdf:jpdf} this is a non-trivial computation for $M\geq2$. In complete analogy to the square case~\cite{AKW:2013}, it turns out that the correlation functions are more easily obtained using a two-matrix model and the method of bi-orthogonal polynomials. We will discuss this in section~\ref{sec:2mm}, including other methods of derivation.

The normalisation constant in equations~\eqref{jpdf:Zs} and~\eqref{jpdf:jpdf} is
\begin{equation}
C_M=N_0!\prod_{n=1}^{N_0}\prod_{m=0}^M\Gamma[n+\nu_m],
\label{jpdf:norm}
\end{equation}
such that the partition function is equal to unity, which is straightforward to check using the Andr\' eief integration formula. The one-point correlation function (or density) is normalised to the number of singular values,
\begin{equation}
\int_0^\infty ds\, R_1^M(s)=N_0,
\label{jpdf:R_norm}
\end{equation}
which becomes evident in the following section.

\section{Two-Matrix Model and Bi-Orthogonal Polynomials}
\label{sec:2mm}

The purpose of this section is to find an explicit expression for the $k$-point correlation functions~\eqref{jpdf:R}. We will follow the idea in~\cite{AKW:2013} and rewrite our problem as a two-matrix model by keeping the integrals over the $s_a^1$'s and $s_a^{M}$'s in Eq.~\eqref{jpdf:Zs2} while integrating over the remaining variables. Within this model we will exploit the method of bi-orthogonal polynomials to achieve our goal. First, we use the identity~\eqref{meijer:meijer-induc} for the Meijer $G$-function to write the partition function~\eqref{jpdf:Zjpdf} with $M\geq 2$ as
\begin{equation}
Z^M=\prod_{a=1}^{N_{0}}\int_0^\infty ds_a \prod_{i=1}^{N_{0}}\int_0^\infty dt_i\,\widetilde P_\jpdf^M(s\,;t\,),
\label{2mm:Z2mm}
\end{equation}
where the joint probability density function is given by
\begin{equation}
\widetilde P_\jpdf^M(s\,;t\,)=C_M\Delta_{N_0}(s)\Delta_{N_0}(t)\olddet_{1\leq k,\ell,\leq N_0}\big[w_\nu^M(s_k,t_\ell)\big],
\label{2mm:jpdf2mm}
\end{equation}
$s_a\equiv s_a^M$ and $t_a\equiv s_a^1$, and the weight function depending on all indices $\nu_m$ collectively denoted by $\nu$ reads
\begin{equation}
w_\nu^M(s,t)=t^{\nu_1-1}e^{-t}\MeijerG{M-1}{0}{0}{M-1}{-}{\nu_M,\,\nu_{M-1},\,\ldots\,,\,\nu_2}{\frac{s}{t}\,}.
\label{2mm:weight}
\end{equation}
The structure of the joint probability density function~\eqref{2mm:jpdf2mm} is similar to that of the two-matrix model discussed in~\cite{EM:1998}. Although the focus in~\cite{EM:1998} is on a multi-matrix model with an Itzykson--Zuber interaction, the argument given is completely general and applies to our situation as well. The $(k,\ell)$-point correlation functions for this two-matrix model are defined as
\begin{multline}
R_{k,\ell}^M(s\,;t\,)=
\frac{(N_0!)^2}{(N_0-k)!(N_0-\ell)!}\\
\times\prod_{a=k+1}^{N_{0}}\int_0^\infty ds_a \prod_{i=\ell+1}^{N_{0}}\int_0^\infty dt_i\,
\widetilde P_\jpdf^M(s;t).
\end{multline}
Obviously, we can obtain the $k$-point correlation functions~\eqref{jpdf:R} by integrating out all $t_i$'s, i.e. setting $\ell=0$.

The benefit of the two-matrix model is that we can exploit the method of bi-orthogonal polynomials as in~\cite{EM:1998}. We choose a family of monic polynomials $q_j^M(t)=t^j+\cdots$ and $p_j^M(s)=s^j+\cdots$, which are bi-orthogonal with respect to the weight~\eqref{2mm:weight},
\begin{equation}
\int_0^\infty ds \int_0^\infty dt\, w_\nu^M(s,t) q_i^M(t)p_j^M(s)=h_j^M\delta_{ij},
\label{2mm:bi-ortho}
\end{equation}
where $h_j^M$ are constants. Furthermore, we introduce the functions $\psi_j^M(t)$ and $\phi_j^M(s)$ defined as integral transforms of the bi-orthogonal polynomials,
\begin{align}
\psi_j^M(t) &\equiv \int_0^\infty ds\, w_\nu^M(s,t) p_j^M(s),\\
\phi_j^M(s) &\equiv \int_0^\infty dt\, w_\nu^M(s,t) q_j^M(t).
\label{2mm:psi-phi}
\end{align}
Note that $\psi_j^M(t)$ and $\phi_j^M(s)$ are not necessarily polynomials. It is evident from the bi-orthogonality of the polynomials~\eqref{2mm:bi-ortho} that we have the orthogonality relations
\begin{equation}
\int_0^\infty dt\, q_i^M(t)\psi_j^M(t)=\int_0^\infty ds\, p_i^M(s)\phi_j^M(s)=h_j^M\delta_{ij}.
\label{2mm:ortho}
\end{equation}
Moreover, it follows from the discussion in~\cite{EM:1998} that the $(k,\ell)$-point correlation functions are given by a determinantal point process
\begin{equation}
R_{k,\ell}^M(s\,;t\,)=\olddet_{\substack{1\leq a,b\leq k \\ 1\leq i,j\leq \ell}}
\begin{bmatrix}
K_{11}^M(s_a,s_b) & K_{12}^M(s_a,t_j) \\
K_{21}^M(t_i,s_b) & K_{22}^M(t_i,t_j)
\end{bmatrix},
\label{2mm:R-kl}
\end{equation}
where the four sub-kernels are defined in terms of the bi-orthogonal polynomials and the weight function as
\begin{align}
K_{11}^M(s_a,s_b)&=\sum_{n=0}^{N_0-1}\frac{p_n^M(s_a)\phi_n^M(s_b)}{h_n^M},                    \nn\\    K_{12}^M(s_a,t_j)&=\sum_{n=0}^{N_0-1}\frac{p_n^M(s_a)q_n^M(t_j)}{h_n^M}, \nn\\
K_{21}^M(t_i,s_b)&=\sum_{n=0}^{N_0-1}\frac{\psi_n^M(t_i)\phi_n^M(s_b)}{h_n^M}-w_\nu^M(s_b,t_i),    \nn\\    K_{22}^M(t_i,t_j)&=\sum_{n=0}^{N_0-1}\frac{\psi_n^M(t_i)q_n^M(t_j)}{h_n^M}.
\label{2mm:kernel}
\end{align}
In particular we have that the $k$-point correlation functions~\eqref{jpdf:R} for the singular values of the product matrix $Y_MY_M^\dagger$ are given by
\begin{equation}
R_{k}^M(s_1,\ldots,s_k)=\olddet_{1\leq a,b\leq k}\big[ K_{11}^M(s_a,s_b) \big].
\label{2mm:R-k}
\end{equation}
The goal is to find the bi-orthogonal polynomials, $q_j^M(t)$ and $p_j^M(s)$, and the norms, $h_j^M$, and thereby all correlation functions for the singular values of the product matrix, $\mY_M$. Note that we use a slightly different notation for the sub-kernels than in~\cite{AKW:2013}; the notation in this paper is chosen to emphasise the fact that all the statistical properties of the singular values are determined by the bi-orthogonal polynomials, $q_j^M(t)$ and $p_j^M(s)$, and the weight function, $w_\nu^M(s,t)$.

In order to find the bi-orthogonal polynomials we follow the approach in~\cite{AKW:2013} and start by computing the bimoments
\begin{equation}
I_{ij}^M\equiv\! \int_0^\infty\!\! ds\! \int_0^\infty\!\! dt\, w_\nu^M(s,t) s^i\,t^j = (i+j+\nu_1)!\!\prod_{m=2}^M(i+\nu_m)!
\label{2mm:bimoment}
\end{equation}
for $M\geq 2$. Here the integration has been performed using integral identities for the Meijer $G$-function, see equations~\eqref{meijer:meijer-mellin} and~\eqref{meijer:meijer-induc}. Using Cramer's rule, the bi-orthogonal polynomials as well as the norms can be expressed in terms of the bimoments as~\cite{BGS:2009a,BGS:2009b},
\begin{align}
q_n^M(t)&=\frac{1}{D_{n-1}^M}
\det\left[
\begin{matrix}
I_{00}^M 	& I_{10}^M 	& \cdots 	& I_{(n-1)0}^M 	& 1      \\
I_{01}^M 	& I_{11}^M 	& \cdots 	& I_{(n-1)1}^M 	& t      \\
\vdots   	& \vdots   	&       	& \vdots       	& \vdots \\
I_{0n}^M 	& I_{1n}^M 	& \cdots 	& I_{(n-1)n}^M 	& t^n
\end{matrix}
\right], \nn\\
p_n^M(s)&=\frac{1}{D_{n-1}^M}
\det\left[
\begin{matrix}
I_{00}^M 	& I_{01}^M 	& \cdots 	& I_{0(n-1)}^M 	& 1      \\
I_{10}^M 	& I_{11}^M 	& \cdots 	& I_{1(n-1)}^M 	& s      \\
\vdots   	& \vdots   	&       	& \vdots       	& \vdots \\
I_{n0}^M 	& I_{n1}^M 	& \cdots 	& I_{n(n-1)}^M 	& s^n
\end{matrix}
\right],
\label{2mm:poly-matrix}
\end{align}
where
\begin{equation}
D_n^M \equiv \olddet_{0\leq i,j\leq n} [I_{ij}^M]=\prod_{i=0}^n\prod_{m=0}^M(i+\nu_m)!.
\end{equation}
The norms can be expressed as
\begin{equation}
h_n^M ={D_n^M}/{D_{n-1}^M}=\prod_{m=0}^M(n+\nu_m)!.
\label{norm}
\end{equation}
Recall that $\nu_i\equiv N_i-N_0\geq 0$ are non-negative integers by definition ($\nu_0=0$).

In order to get more explicit expressions for the bi-orthogonal polynomials, we define the bimoment matrix~\eqref{2mm:bimoment} for $M=1$ as the bimoments with respect to the Laguerre weight,
\begin{equation}
I_{ij}^{M=1}\equiv \int_0^\infty ds\, e^{-s}s^{\nu_1+i+j}=(i+j+\nu_1)!.
\label{2mm:bimoment1}
\end{equation}
It follows that the polynomials~\eqref{2mm:poly-matrix} for $M=1$ are the Laguerre polynomials in monic normalisation,
\begin{equation}
p_n^{M=1}(s)=q_n^{M=1}(s)=\widetilde L_n^{\nu_1}(s)\equiv (-1)^nn!L_n^{\nu_1}(s),
\end{equation}
where $L_n^{\nu_1}(s)$ are the associated Laguerre polynomials. We recall that the Laguerre polynomials are defined as
\begin{equation}
\widetilde L_n^{\nu_1}(s)=\sum_{k=0}^n\frac{(-1)^{n+k}}{(n-k)!}\frac{(n+\nu_1)!}{(k+\nu_1)!}\frac{n!}{k!}s^k
\label{2mm:L-ortho}
\end{equation}
and satisfy the orthogonality relation
\begin{equation}
\int_0^\infty ds\,e^{-s}s^{\nu_1}\widetilde L_k^{\nu_1}(s)\widetilde L_\ell^{\nu_1}(s)=h_k^{M=1}\delta_{k\ell}
\end{equation}
with $h_k^{M=1}=k!(k+\nu_1)!$.

The bimoment matrix, $[I_{ij}^M]_{0\leq i,j\leq n}$, with $M\geq2$ given by equation~\eqref{2mm:bimoment} differs from the bimoment matrix, $[I_{ij}^1]_{0\leq i,j\leq n}$, given by equation~\eqref{2mm:bimoment1} by multiplication of a diagonal matrix. It directly follows from this fact that the polynomials $q_n^M(t)$ are related to the Laguerre polynomials as
\begin{equation}
q_n^M(t)=\prod_{i=0}^{n-1}\prod_{m=2}^M(i+\nu_m)!\frac{D_{n-1}^{1}}{D_{n-1}^M}\widetilde L_n^{\nu_1}(t)=\widetilde L_n^{\nu_1}(t).
\label{2mm:p}
\end{equation}

The evaluation of the polynomials $p_n^M(s)$ is slightly more complicated. For the polynomials $q_n^M(t)$, the factorisation is the same for all powers of $t$, but for the polynomials $p_n^M(s)$ we have to treat the powers differently; in particular we substitute $s^k\to s^k/\prod_{m=2}^M(k+\nu_m)!$. Using the explicit expression for the Laguerre polynomials~\eqref{2mm:L-ortho} we find
\begin{equation}
p_n^M(s)
=\sum_{k=0}^n\frac{(-1)^{n+k}n!}{(n-k)!}\left(\prod_{m=1}^M\frac{(n+\nu_m)!}{(k+\nu_m)!}\right)\frac{s^k}{k!},
\label{2mm:q_poly}
\end{equation}
which is a generalised hypergeometric polynomial (see equation~\eqref{meijer:hyper-poly} in appendix~\ref{sec:meijer})
\begin{equation}
p_n^M(s)=(-1)^n\prod_{m=1}^M\frac{(n+\nu_m)!}{\nu_m!}\\
\hypergeometric{1}{M}{-n}{1+\nu_M,\,\ldots\,,\,1+\nu_1}{s}.
\label{2mm:q_F}
\end{equation}
For $\nu_M=\cdots=\nu_1=0$ this polynomial reduces to the result presented in~\cite{AKW:2013}, while the monic Laguerre polynomials are reobtained by setting $M=1$. Alternatively we may write $p_n^M(s)$ as a Meijer $G$-function,
\begin{equation}
p_n^M(s)=(-1)^n\prod_{m=0}^M(n+\nu_m)!\,\MeijerG{1}{0}{1}{M+1}{n+1}{0,\,-\nu_M,\,\ldots\,,\,-\nu_1}{s}.
\label{2mm:q_G}
\end{equation}
This expression will be particularly useful in section~\ref{sec:moment}, where we discuss the asymptotic behaviour of the endpoints of support of the spectral density. In equation~\eqref{2mm:q_G} we have used the relation~\eqref{meijer:hyper-meijer} between generalised hypergeometric polynomials and Meijer $G$-functions. It might not be immediately clear that the Meijer $G$-function in~\eqref{2mm:q_G} is a polynomial. To see this, one writes the Meijer $G$-function as a contour integral using its definition~\eqref{meijer:meijer-def}. The integrand has exactly $n$ simple poles and the contour is closed such that these poles are encircled. The residue for each pole gives a monomial, such that the complete contour integral yields a polynomial.

With the explicit expressions for the bi-orthogonal polynomials~\eqref{2mm:p} and~\eqref{2mm:q_F}, we are ready to compute the functions $\psi_n^M(t)$ and $\phi_n^M(s)$ defined in equation~\eqref{2mm:psi-phi}, and thereby implicitly find all the sub-kernels~\eqref{2mm:kernel}. The functions $\psi_n^M(t)$ turn out to be polynomials, too,
\begin{equation}
\psi_n^M(t)=\prod_{m=2}^M(n+\nu_m)!\,t\widetilde L_n^{\nu_1}(t),
\label{2mm:psi}
\end{equation}
which can be directly obtained from the definition~\eqref{2mm:psi-phi} using the integral identity~\eqref{meijer:meijer-mellin}.

Likewise, we can obtain an explicit expression for the functions $\phi_n^M(s)$ by inserting the polynomial~\eqref{2mm:p} into the definition~\eqref{2mm:psi-phi}. It follows from the integral identity~\eqref{meijer:meijer-induc} that
\begin{multline}
\phi_n^M(s)=\sum_{k=0}^n\frac{(-1)^{n+k}}{(n-k)!}\frac{(n+\nu_1)!}{(k+\nu_1)!}\frac{n!}{k!}\\
\times\MeijerG{M}{0}{0}{M}{-}{\nu_M,\,\ldots\,,\,\nu_2,\nu_1+k}{s}.
\label{2mm:phi-long}
\end{multline}
However, it is possible to get a more compact expression. Recall that the Laguerre polynomials can be expressed using Rodrigues' formula,
\begin{equation}\label{LagRod}
\widetilde L_n^{\nu_1}(t)=(-1)^nt^{-\nu_1}e^t\frac{d^n}{dt^n}\big(t^{n+\nu_1}e^{-t}\big).
\end{equation}
We insert Rodrigues' formula into the definition for $\phi_n^M(s)$, see equation~\eqref{2mm:psi-phi}. The differentiation in equation~\eqref{LagRod} can easily be changed to a differentiation of the Meijer $G$-function (stemming from the weight function) using integration by parts, since all boundary terms are zero. Then the differentiation can be computed using equation~\eqref{meijer:meijer-diff}, while the final integration over $t$ can be performed using the identity~\eqref{meijer:meijer-induc}. This finally leads to
\begin{equation}
\phi_n^M(s)=(-1)^n\MeijerG{M}{1}{1}{M+1}{-n}{\nu_M,\,\nu_{M-1},\,\ldots\,,\,\nu_1,0}{s}.
\label{2mm:phi}
\end{equation}
In addition to the fact that equation~\eqref{2mm:phi} is a more compact expression than the representation~\eqref{2mm:phi-long}, we is also immediate that $\phi_n^M(s)$ is symmetric in all the indices $\nu_m$, which is far from obvious in equation~\eqref{2mm:phi-long}.

Now we have explicit expressions for all components contained in the formula for the $(k,\ell)$-point correlation functions~\eqref{2mm:R-kl}, which completes the derivation. In particular combining equations~\eqref{norm},~\eqref{2mm:q_F},  and~\eqref{2mm:phi} the sub-kernel $K_{11}^M(s_a,s_b)$ is  given by
\begin{multline}
K_{11}^M(s_a,s_b)=
\!\sum_{n=0}^{N_0-1}\!\!\frac{1}{n!}\!\prod_{m=1}^M\frac{1}{\nu_m!}
\hypergeometric{1}{M}{-n}{1+\nu_M,\,\ldots\,,\,1+\nu_1}{s_a} \\
\times\MeijerG{M}{1}{1}{M+1}{-n}{\nu_M,\,\ldots\,,\,\nu_1,0}{s_b}.
\label{2mm:kernel11_F}
\end{multline}
It provides a direct generalisation of the formula given in~\cite{AKW:2013} for square matrices to the case of rectangular matrices. If we use the alternative formula~\eqref{2mm:q_G} for $p_n^M(s)$ we obtain
\begin{align}
K_{11}^M(s_a,s_b)=
\sum_{n=0}^{N_0-1}&\MeijerG{1}{0}{1}{M+1}{n+1}{0,\,-\nu_M,\,\ldots\,,\,-\nu_1}{s_a} \nn\\
&\times\MeijerG{M}{1}{1}{M+1}{-n}{\nu_M,\,\ldots\,,\,\nu_1,0}{s_b}.
\label{2mm:kernel11_G}
\end{align}
The $k$-point correlation functions for the singular values are immediately found from equation~\eqref{2mm:R-k}. Note that the kernel and thereby all $k$-point correlation functions are symmetric in all the indices $\nu_m$. This symmetry reflects the invariance of the singular values of the product matrix, $Y_M=X_M\cdots X_1$, under reordering of the matrices $X_m$ which we prove in a more general setting in \cite{IK:2013}.
The normalisation of the spectral density~\eqref{jpdf:R_norm} is immediately clear from the orthogonality relation~\eqref{2mm:ortho}.

Finally we would like to mention an alternative derivation for the correlation functions~\eqref {jpdf:R} in terms of the kernel $K_{11}^M$. 
Given the orthogonality relation~\eqref{2mm:ortho} of the polynomials $p_i^M$~\eqref{2mm:q_poly} and the functions
$\phi_j^M$~\eqref{2mm:phi-long} we can generate these by adding columns in the two determinants in the joint probability density function~\eqref{jpdf:jpdf} and then proceed with the standard Dyson theorem. This is in complete analogy as described in~\cite{AKW:2013}. 
Alternatively, the kernel can be derived by using bi-orthogonal functions and explicitly inverting the bimoment matrix~\cite{ES:2013}.
Furthermore, a construction using multiple orthogonal polynomials exist~\cite{Zhang:2013,KZ:2013}, too.

\section{Moments and Asymptotics}
\label{sec:moment}

In this section we take a closer look at the spectral density. First we will use the density to find an explicit expression for the moments. Second we will discuss the macroscopic large-$N_0$ limit of the density. 

\begin{figure}[htbp]
\centering
\includegraphics{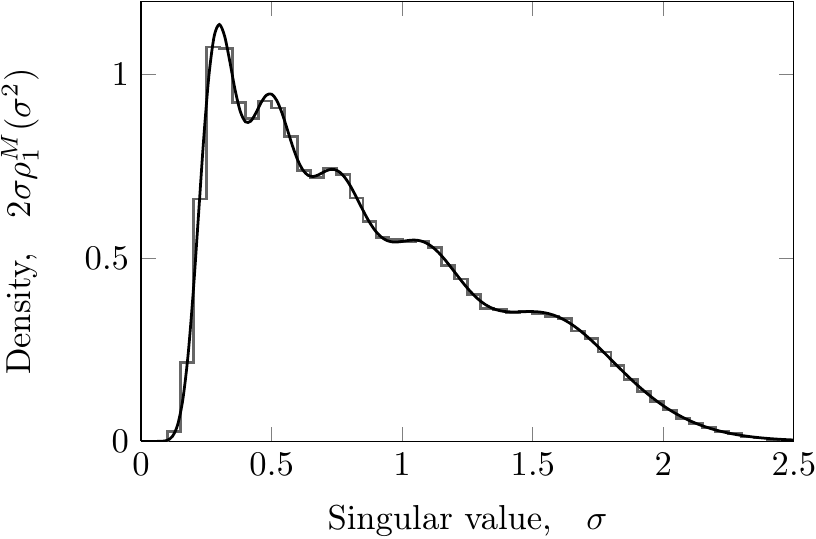}\par
\vspace*{3mm}
\includegraphics{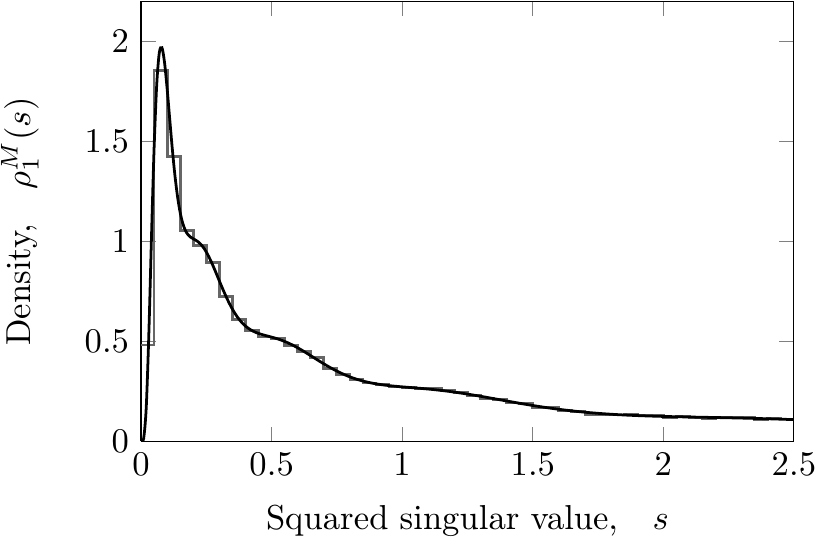}
\caption{The histograms (bin width is $0.05$) show the distributions of singular values (top) and squared singular values (bottom) for $50\,000$ realisations of the product matrix $\mY_3=\mX_3\mX_2\mX_1$ for $M=3$, with $\nu_1=5$, $\nu_2=10$, $\nu_3=15$ and $N_0=5$. The solid curves are the analytical predictions for the rescaled densities of singular values, $2\sigma \rho_{1}^{3}(\hat\sigma^2)$, and of squared singular values, $\rho_{1}^{3}(\hat s)$, respectively, cf. equation~\eqref{moment:density-scaled}. 
}
\label{fig:density}
\end{figure}

We know from the previous section that the density, or one-point correlation function, is given as a sum over Meijer $G$-functions,
\begin{align}
R_{1}^M(s)=
\sum_{n=0}^{N_0-1}&\MeijerG{1}{0}{1}{M+1}{n+1}{0,\,-\nu_M,\,\ldots\,,\,-\nu_1}{s} \nn\\
&\times\MeijerG{M}{1}{1}{M+1}{-n}{\nu_M,\,\ldots\,,\,\nu_1,0}{s},
\label{moment:density}
\end{align}
which is normalised to the number of singular values, $N_0$. Figure~\ref{fig:density} shows a comparison between the analytical expression and numerical simulations for an example. The expectation value for the singular values is defined in terms of the density~\eqref{moment:density} as
\begin{equation}
\E\{f(s)\}\equiv \frac{1}{N_0}\int_0^\infty ds\,R_1^M(s)\,f(s),
\label{moment:expect}
\end{equation}
where the factor $1/N_0$ is included since the density~\eqref{moment:density} is normalised to the number of singular values.

We will first look at the moments, $\E\{s^\ell\}$. Note that we do not assume that $\ell$ is an integer, and that the half-integer values of $\ell$ are interesting, too, since the singular values, $\sigma_a$, of the product matrix, $\mY_M$, are given by the square roots of the eigenvalues of the Wishart matrix, i.e. $\sigma_a=\sqrt{s_a}$. In order to calculate the moments, we  explicitly write the first Meijer $G$-function in equation~\eqref{moment:density} as a polynomial, see equations~\eqref{2mm:q_poly} and~\eqref{2mm:q_G}, and rewrite the moments as
\begin{align}
N_0\E\{s^\ell\}&=\sum_{n=0}^{N_0-1}\sum_{k=0}^n\frac{(-1)^k}{(n-k)!}\prod_{m=0}^M\frac{1}{(k+\nu_m)!} \nn\\
&\times\int_0^\infty ds\,s^{\ell+k}\MeijerG{M}{1}{1}{M+1}{-n}{\nu_M,\,\ldots\,,\,\nu_1,0}{s}.
\end{align}
The integral over $s$ can be performed using an identity for the Meijer $G$-function~\eqref{meijer:meijer-mellin}. After reordering the sums and applying Euler's reflection formula for the gamma-function we get
\begin{multline}
N_0\E\{s^\ell\}=\sum_{k=0}^{N_0-1}\prod_{m=0}^M\frac{\Gamma[\ell+k+\nu_m+1]}{(k+\nu_m)!} \\
\times\sum_{n=0}^{N_0-k-1}\frac{(-1)^n}{n!\,\Gamma[\ell-n+1]},
\label{moment:moment-sum-sum}
\end{multline}
where $\ell$ may also take non-integer values.
For integer values of $\ell$ some of the terms will vanish due to the poles of the gamma-function.
Note that the moments are divergent whenever $\ell\leq -\nu_{\text{min}}-1$ is an integer ($\nu_\text{min}\equiv\min\{\nu_1,\ldots,\nu_M\}$), but well-defined for all other values of $\ell$.
The second sum in equation~\eqref{moment:moment-sum-sum} can be evaluated by a relation for the (generalised) binomial series 
\begin{equation}
\sum_{n=0}^N (-1)^n\begin{pmatrix} z \\ n \end{pmatrix}=(-1)^N\begin{pmatrix} z-1 \\ N \end{pmatrix},\quad z\in\C.
\label{moment:binomial}
\end{equation}
We write the first sum in equation~\eqref{moment:moment-sum-sum} in reverse order ($k\to N_0-k-1$) and perform the second sum using the identity~\eqref{moment:binomial} yielding
\begin{eqnarray}\label{moment:moment}
N_0\E\{s^\ell\}&=&\sum_{k=0}^{N_0-1}\frac{(-1)^k}{k!\Gamma[\ell-k]\ell}\prod_{m=0}^M\frac{\Gamma[\ell+N_m-k]}{\Gamma[N_m-k]}.
\end{eqnarray}
Alternatively, the moments can be written as
\begin{eqnarray}
N_0\E\{s^\ell\}&=&\sum_{k=0}^{N_0-1}\frac{(-1)^{1+k}\prod_{j=0}^{N_0-1}(j-\ell-k)}{k!(N_0-1-k)!\ell}
\label{moment:moment.1}\\
&&\qquad\times\prod_{m=1}^M\frac{\Gamma[\ell+\nu_m+k+1]}{\Gamma[\nu_m+k+1]}\nonumber
\end{eqnarray}
which is useful when considering the limit of negative integer $\ell$.
Recall that $N_m$ are the different matrix dimensions of the original product~\eqref{intro:product} and $\nu_m=N_m-N_0$.

For $\ell\to0$ all terms in the sum are equal to one and we recover the normalisation. Simplifications also occur when $\ell$ is an integer; here most of the terms in the sum vanish, due to the gamma-function in the denominator. In particular, the first positive moment and the first negative moment are given by
\begin{equation}
\E\{s\}\equiv \cN_M=\prod_{m=1}^MN_m
\ \text{ and }\ 
\E\{s^{-1}\}=\prod_{m=1}^M\frac{1}{\nu_m}.
\end{equation}
The second moment is slightly more complicated,
\begin{equation}
\E\{s^2\}=\frac{1}{2}\prod_{m=1}^MN_m\bigg[\prod_{m=0}^M(N_m+1)-\prod_{m=0}^M(N_m-1)\bigg].
\end{equation}
When $M=1$ these formulae reduce to the well-known results for the Wishart--Laguerre ensemble (e.g. see~\cite{TV:2004}), while we get the result~\cite{AKW:2013} for square matrices by setting $N_0=\cdots=N_M$. Note that any negative moment is divergent if $\nu_m=0$ for any $1\leq m\leq M$.

The first moment, $\cN_M$, provides us with a natural scaling of the spectral density,
\begin{equation}
\rho_1^M(\hat s)\equiv \frac{\cN_M}{N_0}R_1^M(\hat s\,\cN_M),
\label{moment:density-scaled}
\end{equation}
such that the rescaled density has a finite first moment of unity also in the large-$N_0$ limit. In equation~\eqref{moment:density-scaled} and the following, we use a hat `\,$\widehat{\ }$\,' to denote rescaled variables. 

The expectation value with respect to the rescaled density~\eqref{moment:density-scaled} is related to the definition~\eqref{moment:expect} by a simple scaling of the variable,
\begin{equation}
\hat \E\{f(\hat s)\}\equiv\int_0^\infty d\hat s\, \rho_1^M(\hat s)f(\hat s)=\E\left\{f\left(\frac{\hat s}{\cN_M}\right)\right\},
\end{equation}
for any observable $f(\hat s)$. The rescaling ensures that we have a well-defined probability density with compact support in the large-$N_0$ limit; in particular the density $\rho_1^{1}(\hat s)$ for a single matrix $M=1$ reduces to the celebrated Mar\v cenko--Pastur density for $N_0\to\infty$.

An algebraic way to obtain the macroscopic behaviour of the spectral density~\eqref{moment:density-scaled} for arbitrary $M$ was provided in~\cite{BJLNS:2010}, using the resolvent also known as the Stieltjes transform, $G^M(\hat z)$,  defined as 
\begin{equation}
G^M(\hat z)\equiv\int_0^\infty d \hat{s} \lim_{N_0\to\infty}\frac{
\rho_1^M(\hat s)}{\hat{z}-\hat{s}},
\end{equation}
with $\hat z$ outside the limiting support of $\rho_1^M$. It was shown that in the large-$N_0$ limit the resolvent satisfies a polynomial equation~\cite{BJLNS:2010},
\begin{equation}
\hat z\,G^M(\hat z)\prod_{m=1}^M\frac{\hat z\,G^M(\hat z)+\hat\nu_m}{\hat\nu_m+1}=\hat z(\hat z\,G^M(\hat z)-1),
\label{moment:resolvent}
\end{equation}
where $\hat z$ lies outside the support of the singular values and $\hat \nu_m$ denotes the rescaled differences in matrix dimensions, i.e. $\hat \nu_m\equiv\nu_m/N_0$ for $m=1,\ldots,M$. In general one needs to solve an $(M+1)$-st order equation in order to find the resolvent, $G^M(\hat z)$. It is clear, that such an equation can generically only be solved analytically for $M\leq 3$ (see also the discussions in~\cite{Zhang:2013,PZ:2011}).

\begin{figure}[htpb]
\centering
\includegraphics{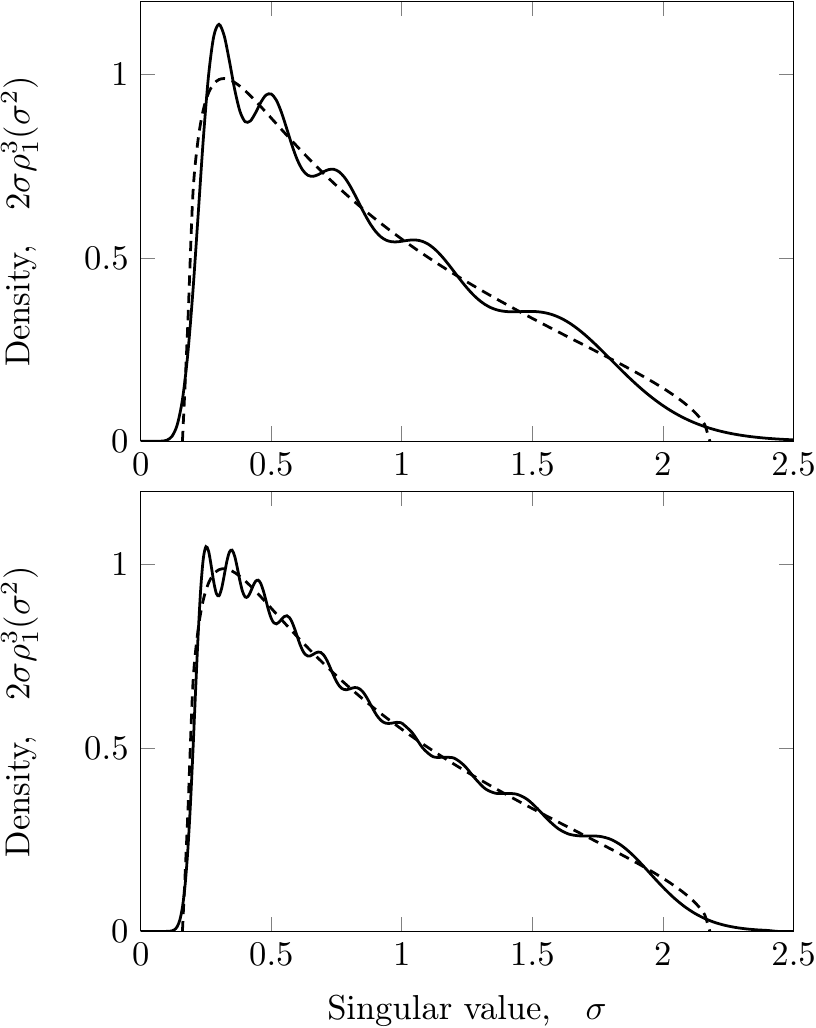}
\caption{The solid lines show the $M=3$ rescaled spectral densities for the singular values for $N_0=5$ (top) and $N_0=10$ (bottom) both with $\hat\nu_1=1$, $\hat\nu_2=2$, $\hat\nu_3=3$. The dashed curves indicate the corresponding macroscopic limit~\cite{BJLNS:2010}.}
\label{fig:macro}
\end{figure}

The correct resolvent is chosen by its asymptotic behaviour, $\hat zG^M(\hat z)\to 1$ for $\hat z\to\infty$. When an expression for the resolvent is known, then the spectral density can be directly obtained from the resolvent using
\begin{equation}
\rho_1^{M,\infty}(\hat s)\equiv\lim_{N_0\to\infty}\rho_1^M(\hat s)=\frac{1}{\pi}\lim_{\epsilon\to 0^+} \Im G^M(\hat s-\imath\epsilon).
\end{equation}
In figure~\ref{fig:macro} we compare this macroscopic limit with the rescaled density~\eqref{moment:density} at finite $N_m$.

For the case $M=1$ one can readily derive the well-known Mar\v cenko--Pastur law. Another particular case in which the spectral density $\rho_1^{M,\infty}$ can be directly calculated is  $M=2$ with $\hat{\nu}_{1}$ and $\hat{\nu}_{2}$ arbitrary. This case plays an important role when studying cross correlation matrices of two different sets of time series as it appears in forecasting models \cite{Bouchaud,Vin13} where time-lagged correlation matrices are non-symmetric. Our random matrix model then corresponds to the case of two time series which are uncorrelated. Despite the independence of the distribution of the matrix elements correlations among the singular values of the cross correlation matrix follow. The solution of equation~\eqref{moment:resolvent} yields the level density
\begin{eqnarray}
 \rho^{M,\infty}_1(\hat{s})&=&\frac{\sqrt{3(\hat\nu_1+1)(\hat\nu_2+1)\hat{s}+\hat\nu_1^2-\hat\nu_1\hat\nu_2+\hat\nu_2^2}}{3\pi\hat{s}}\nonumber\\
 &&\times\Im\left[A^{-1/3}(f(\hat{s}))+A^{1/3}(f(\hat{s}))\right]\label{M2-level}
\end{eqnarray}
with
\begin{eqnarray}\label{def-f}
 &&f\left(\frac{z}{(\hat\nu_1+1)(\hat\nu_2+1)}\right)\\
 &=&3\frac{\left[3z+\hat\nu_1^2-\hat\nu_1\hat\nu_2+\hat\nu_2^2\right]^3}{\left[3(3+\hat\nu_1+\hat\nu_2)z+\hat\nu_1^3-(\hat\nu_1+\hat\nu_2)^3/3+\hat\nu_2^3\right]^{2}}\nonumber
\end{eqnarray}
and
\begin{equation}\label{def-A}
 A(z)=\sqrt{\frac{27}{4 z}-1}-\sqrt{\frac{27}{4 z}}.
\end{equation}
Indeed the special case $\hat\nu_1=\hat\nu_2=0$ agrees with the result derived in \cite{ZPNC:2011,AKW:2013,Zhang:2013} because $f(\hat{s})|_{\hat\nu_1=\hat\nu_2=0}=\hat{s}$.

It is also desirable to know where the endpoints of support of the macroscopic spectrum are located. These edges can be found from the algebraic formula for the resolvent~\eqref{moment:resolvent} using a simple trick. We assume that the resolvent behaves as $\abs{G^M(\hat z)}\sim \abs{\hat z-\hat s_\pm}^{\alpha_\pm}$ with $\alpha_\pm<1$  and  $\alpha_\pm\neq0$ in the vicinity of the edges, $\hat s_\pm$. This edge behaviour of the resolvent is known to hold in certain cases, e.g. $M=1$ yields $\alpha_\pm=1/2<1$ (except when the inner edge is zero, $\hat s_-=0$, then $\alpha_-=-1/2<1$). Due to known universality results for random matrices, it is expected that $\alpha_\pm<1$ and $\alpha_\pm\neq0$ in general. With this particular edge behaviour, it is clear that $\abs{dG^M/d\hat z}\to\infty$ for $\hat z\to\hat s_\pm$, or equivalently $d\hat z/dG^M\to 0$ for $\hat z\to\hat s_\pm$. Differentiating both sides of equation~\eqref{moment:resolvent} with respect to $G^M$ and evaluating them at $d\hat z/dG^M=0$ yields an equation for the extrema of $\hat z$,  
\begin{equation}
\hat z_0=\bigg(1+\sum_{j=1}^M\frac{\hat z_0 G^M(\hat z_0)}{\hat z_0 G^M(\hat z_0)+\hat\nu_j} \bigg)\prod_{m=1}^M\frac{\hat z_0 G^M(\hat z_0)+\hat\nu_m}{\hat\nu_m+1}.
\end{equation}
Two of these extrema are the inner edge, $\hat z_0=\hat s_-$, and the outer edge, $\hat z_0=\hat s_+$.
The edges, $\hat s_\pm$, also satisfy equation~\eqref{moment:resolvent}. Combining both equations, we get an expression for the edges
\begin{equation}
\hat s_\pm=\frac{\hat u_0}{1+\hat u_0}\prod_{m=1}^M\frac{\hat\nu_m-\hat u_0}{\hat\nu_m+1},
\label{moment:edge-eq1}
\end{equation}
in terms of $\hat u_0\equiv -\hat z_0G^M(\hat z_0)$ which is given by
\begin{equation}
\sum_{m=1}^M\frac{\hat u_0(\hat u_0+1)}{\hat\nu_m-\hat u_0}=1.
\label{moment:edge-eq2}
\end{equation}
This equation is equivalent to a polynomial equation of $(M+1)$'st order as it is the case for the resolvent, see equation~\eqref{moment:resolvent}. However, in certain cases equation~\eqref{moment:edge-eq2} simplifies. In particular, equation~\eqref{moment:edge-eq2} reduces to an $M$-th order equation if $\hat\nu_i=\hat\nu_j$ for $i\neq j$, if $\hat\nu_i\to0$ or if $\hat\nu_i\to\infty$. The latter means that $N_i\gg N_0$ meaning that the matrix dimension $N_i$ decouples from the macroscopic theory.

\begin{figure}[htpb]
\centering
\includegraphics{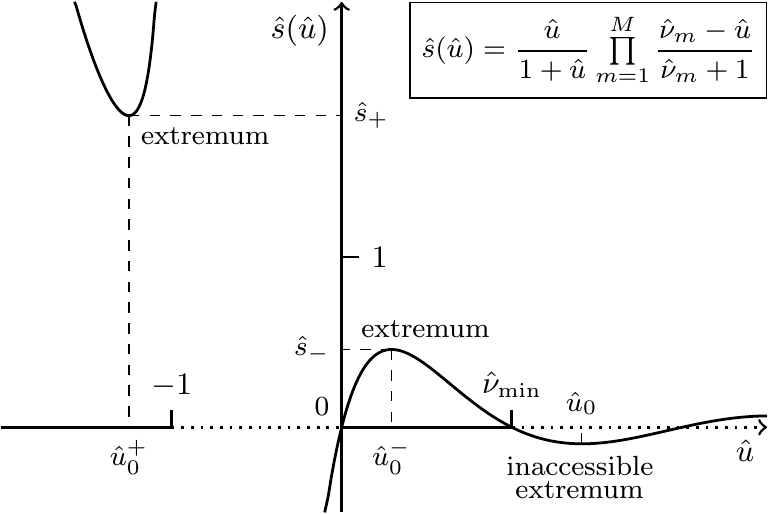}
\caption{Illustration of the optimisation problem given by equation~\eqref{moment:saddle2}. Extrema within the intervals $(-1,0)$ and $(\hat\nu_\text{min},\infty)$ must be disregarded due to the cuts in the complex ($-\hat v$)-plane and complex $\hat u$-plane, respectively. This leaves only two valid extrema which correspond to the inner edge and the outer edge, respectively. Note that the solutions for the inner edge and the outer edge are separated by the pole at $-1$.}
\label{fig:saddle}
\end{figure}

In general the set of equations~(\ref{moment:edge-eq1}) and (\ref{moment:edge-eq2}) yields $(M+1)$ solutions of which two correspond to the inner and outer edge of the spectral density. In the special case where $\hat\nu\equiv\hat\nu_1=\cdots=\hat\nu_M$, there are only two solutions (see figure~\ref{fig:saddle})
\begin{align}
\hat s_\pm(\hat{\nu}) =\,{} &\frac{M+1+2\hat\nu\pm\sqrt{(M+1)^2+4M\hat\nu}}{2(\hat\nu+1)} \nn\\
&\times\left(\frac{M+1+2M\hat\nu\pm\sqrt{(M+1)^2+4M\hat\nu}}{2M+2M\hat\nu}\right)^M.\label{resdegnu}
\end{align}
Note that for $M=1$ this result reduces to the known values for the edges of the Mar\v cenko--Pastur density (e.g. see~\cite{TV:2004}), while the limit $\hat{s}_\pm(\hat\nu\to0)$ reproduces the result for the product of square matrices, see~\cite{ZPNC:2011,AKW:2013,Zhang:2013}. It is easy to numerically verify that the result holds in general.

Looking at the equations~\eqref{moment:edge-eq1} and~\eqref{moment:edge-eq2}, an obvious question is: Which solutions correspond to the edges of the spectrum? In order to answer this question, we will derive the same equations through a different route. The rescaled spectral density~\eqref{moment:density-scaled} serves as the starting point, and the locations of the edges are determined using a saddle point approximation for large $N_0$. This also illustrates the point that the finite $N_m$ expression discussed in this paper is equivalent to the result presented in~\cite{BJLNS:2010} in the macroscopic limit. 

In the large-$N_0$ limit we may approximate the sum over $n$, see equation~\eqref{moment:density}, by an integral. Moreover, we write the Meijer $G$-functions as contour integrals~\eqref{meijer:meijer-def} and approximate the gamma-functions using Stirling's formula. The rescaled density~\eqref{moment:density-scaled} becomes
\begin{align}
\rho_{1}^M(\hat s)\approx \frac{\cN_M}{N_0}\int_0^1 d\hat n\,
&\frac{N_0}{2\pi \imath}\int_{L_1} d\hat v\, e^{-N_0S(-\hat v,\hat n)} \nn\\
\times &\frac{N_0}{2\pi \imath}\int_{L_2} d\hat u\, e^{N_0S(\hat u,\hat n)},
\label{moment:density-approx}
\end{align}
where the action, $S$, is given by
\begin{align}
S(\hat u,\hat n)=\hat{u}{\rm ln}\cN_M\hat s &+\sum_{m=1}^M(\hat\nu_m-\hat u)({\rm ln} N_0(\hat \nu_m-\hat u)-1) \nn \\
&+(\hat n+\hat u)({\rm ln} N_0(\hat n+\hat u)-1) \nn \\
&-\hat u\,({\rm ln} N_0\hat u-1)
\end{align}
with $\hat n=n/N_0$, $\hat u=u/N_0$ and $\hat\nu_m=\nu_m/N_0$. It is important to note that the integrand in the definition of the Meijer $G$-function~\eqref{meijer:meijer-def} contains poles which lie on the real axis. The contours $L_1$ and $L_2$ encircle the poles of the original Meijer $G$-functions in accordance to definition~\eqref{meijer:meijer-def}. In the large-$N_0$ limit these poles condense into cuts, such that the complex $\hat u$-plane has a cut on the interval $(\hat\nu_{\text{min}},\infty)$ and the complex $(-\hat v)$-plane has a cut on the interval $(-1,0)$. The contours $L_1$ and $L_2$ encircle these cuts in the $\hat v$-plane  and the $\hat u$-plane, respectively. Both contour integrals can be evaluated by a saddle point approximation. Furthermore, variation with respect to $\hat n$ yields $\hat u=-\hat v$ at the saddle point and due to the symmetry between the two saddle point equations we can restrict our attention to one of them. The saddle point equation for $\hat u$ yields
\begin{equation}
\hat s=\frac{\hat u_0}{\hat n+\hat u_0}\prod_{m=1}^M\frac{\hat\nu_m-\hat u_0}{\hat\nu_m+1},\qquad 0\leq \hat n\leq 1.
\label{moment:saddle}
\end{equation}
Equation~\eqref{moment:saddle} gives the saddle points, $\hat u_0$, for any given $\hat s$. In order to find the saddle points for the edges of the spectrum, we have to find the values of $\hat n$ and $\hat u_0$ which give the extremal values of $\hat s$. 

Optimising with respect to $\hat n$, we see that $\hat n$ has no optimal value within the interval $(0,1)$, hence $\hat n$ must lie on the boundary due to the Laplace approximation (saddle point approximation on a real support). The only non-trivial result comes from $\hat n=1$. Inserting this condition into the saddle point equation~\eqref{moment:saddle} we reproduce formula~\eqref{moment:edge-eq1}. The condition for $\hat u_0$ is given by differentiating the left hand side of the saddle point equation~\eqref{moment:saddle} and setting this result equal to zero,
\begin{equation}
\frac{d}{d\hat u_0}\left[\frac{\hat u_0}{1+\hat u_0}\prod_{m=1}^M\frac{\hat\nu_m-\hat u_0}{\hat\nu_m+1}\right]=0.
\label{moment:saddle2}
\end{equation}
This condition is identical to formula~\eqref{moment:edge-eq2}. Hence the saddle point method reproduces the result obtained from the algebraic equation~\eqref{moment:resolvent} for the resolvent. 

The saddle points, which satisfy equation~\eqref{moment:saddle2}, are the extrema of the function within the square brackets. This function has a pole at $-1$ and goes to $+\infty$ for $\hat u_0\to -\infty$ such that there is exactly one minimum to the left of the pole, see figure~\ref{fig:saddle}. On the right of the pole the function oscillates such that it has zeros at $0,\hat\nu_1,\ldots,\hat\nu_M$. Since the rational function on the right hand side of equation~\eqref{moment:saddle} is continuous it has extrema between neighbouring zeros, see figure~\ref{fig:saddle}, yielding $M$ additional extrema. It follows that the optimisation problem~\eqref{moment:saddle2} has $M+1$ solutions for $\hat u_0$, which are all real: One solution $\hat u_0^+<-1$ which gives the outer edge of the spectrum $\hat{s}_+$, one solution $0\leq\hat u_0^-\leq\hat\nu_\text{min}$ which gives the inner edge of the spectrum  $\hat{s}_-$, and $M-1$ solutions $\hat u_0\geq\nu_\text{min}$ which must be disregarded due to the cut in the complex $\hat u$-plane mentioned above. It is clear that equation~\eqref{moment:saddle2} cannot have more than $M+1$ solutions implying that we have 
found all solutions. With this result we know how to choose the correct solution of  equation~\eqref{moment:edge-eq2}, which was what we wanted to establish.

Before ending the discussion about the edges of the spectral density, it is worth noting that equation~\eqref{moment:edge-eq2} is an $(M+1)$-st order equation, and the general case can for this reason not be solved analytically. However, it is possible to set up some analytical bounds for the edges. The starting point are the conditions $0\leq \hat u_0^-\leq\hat \nu_\text{min}$ and $-\infty<\hat u_0^+<-1$ for the saddle points. We will analyse step by step first the bounds on the inner edge, $\hat{s}_-$, and then on the outer edge, $\hat{s}_+$. 

Let us consider the inner edge, $\hat{s}_-$. Since $0\leq\hat{\nu}_{\min}\leq\hat{\nu}_m$, $m=1,\ldots,M$, we can readily estimate
\begin{equation}\label{est.1}
\min\left\{\frac{\hat{\nu}_m}{\hat{\nu}_m+1},\frac{\hat{\nu}_{\max}-\hat{u}_0}{\hat{\nu}_{\max}+1}\right\}\geq\frac{\hat{\nu}_m-\hat{u}_0}{\hat{\nu}_m+1}\geq \frac{\hat{\nu}_{\min}-\hat{u}_0}{\hat{\nu}_{\min}+1}
\end{equation}
for any $\hat{u}_0\geq 0$. Note that these bounds hold since the rational function,  $(\hat{\nu}_m-\hat{u}_0)/(\hat{\nu}_m+1)$, is strictly monotonously increasing in $\hat{\nu}_m$ for $\hat{u}_0\geq 0$. We plug equation~\eqref{est.1} into equation~\eqref{moment:edge-eq1} and extremise the lower and upper bound which yields
\begin{equation}
0\leq \hat s_-(\hat{\nu}_{\min})\leq\hat s_-\leq \min\left\{\prod_{m=1}^M\frac{\hat\nu_m}{\hat\nu_m+1},\hat{s}_-(\hat{\nu}_{\max})\right\}<1,
\label{moment:bound-inner}
\end{equation}
where we made use of the result~\eqref{resdegnu} for the case when all $\hat{\nu}$ are equal to $\hat{\nu}_{\min}$ or to $\hat{\nu}_{\max}$. The bounds~\eqref{moment:bound-inner} are not at all optimal. However they immediately reflect the fact that the inner edge vanishes if and only if $\hat \nu_{\min}$ vanishes.

For the outer edge we have to employ the condition $\hat{u}_0<-1$ which yields the estimates
\begin{equation}\label{est.2}
\frac{\hat{\nu}_{\min}-\hat u_0}{\hat{\nu}_{\min}+1}\geq\frac{\hat{\nu}_m-\hat{u}_0}{\hat{\nu}_m+1}\geq \frac{\hat{\nu}_{\max}-\hat{u}_0}{\hat{\nu}_{\max}+1}.
\end{equation}
Hereby we used the fact that the rational function,  $(\hat{\nu}_m-\hat{u}_0)/(\hat{\nu}_m+1)$, is monotonously decreasing in $\hat{\nu}_m$ in the considered regime. Employing the result~\eqref{resdegnu} we find the bounds
\begin{equation}
1<\hat s_+(\hat{\nu}_{\max})\leq\hat s_+\leq\hat s_+(\hat{\nu}_{\min})\leq\frac{(M+1)^{M+1}}{M^M}<\infty.
\label{moment:bound-outer}
\end{equation}
Again the bounds can certainly be improved but they give a good picture what the relation is between the case of degenerate $\hat{\nu}$, cf. equation~\eqref{resdegnu}, and the general case, $\hat\nu_j\neq \hat\nu_i$ for $j\neq i$.

\section{Mutual Information for Progressive Scattering}
\label{sec:mimo}

We will now turn to a brief discussion of the mutual information, which is an important quantity in wireless telecommunication. We look at a MIMO communication channel with multi-fold scattering as mentioned in section~\ref{sec:intro}. The communication link is described by a channel matrix given by a product of complex ($\beta=2$) matrices from the Wishart ensemble as in equation~\eqref{intro:product}. The mutual information is defined as
\begin{align}
\cI(\gamma,s)&=\log_2\det\Big[{\mathbf 1}_{N_0}+\gamma\frac{\mY_M\mY_M^\dagger}{\cN_M}\Big] \nn\\
&=\sum_{a=1}^{N_0}\log_2 \Big(1+\gamma\frac{s_a}{\cN_M} \Big),
\end{align}
where $\gamma$ is the constant signal-to-noise ratio at the transmitter and $s_a$ are the singular values distributed according to the density~\eqref{moment:density}. The mutual information measures an upper bound for the spectral efficiency in bits per time per bandwidth ($\text{bit}/\text{s}/\text{Hz}$).

In order to evaluate the expectation value of the mutual information, the so-called ergodic mutual information, we rewrite the logarithm as a Meijer $G$-function, see equation~\eqref{meijer:meijer-func}. We use the expression~\eqref{2mm:phi-long} for the functions $\phi_n^M(s)$, while we write $p_n^M(s)$ in polynomial form~\eqref{2mm:q_poly}. The integration over the product of two Meijer $G$-functions can be performed using equation~\eqref{meijer:meijer-product}, which finally yields
\begin{align}
\hat\E\{\cI(\gamma,\hat s)\}&=\frac{1}{{\rm ln} 2}\sum_{n=0}^{N_0-1}\sum_{k,\ell=0}^n\frac{(-1)^{k+\ell}}{(n-k)!(n-\ell)!} \nn \\
&\times\frac{n!}{k!\ell!}\frac{(n+\nu_1)!}{(\ell+\nu_1)!}\prod_{m=1}^M\frac{1}{(k+\nu_m)!} \nn\\
&\times\MeijerG{M+2}{1}{2}{M+2}{0,\,1}{k+1+\nu_M,\,\ldots\,,\,k+\ell+1+\nu_1,\,0,\,0}{\gamma^{-1}}\label{mutinf}
\end{align}
For square matrices, i.e. $\nu_i=0$ for all $i=1\ldots M$, this triple sum was derived in \cite{AKW:2013}. Although it is not obvious from this formulation, the mutual information is also independent of the ordering of $\nu_m$. This is reflected after simplifying the expression~\eqref{mutinf} with help of a combination of the equations~\eqref{2mm:L-ortho}, \eqref{LagRod}, \eqref{meijer:meijer-induc}, and \eqref{meijer:meijer-diff} to
\begin{align}
\hat\E\{\cI(\gamma,\hat s)\}&=\frac{1}{{\rm ln} 2}\sum_{n=0}^{N_0-1}\sum_{k=0}^n\frac{(-1)^{k}}{(n-k)!k!}\prod_{m=1}^M\frac{1}{(k+\nu_m)!} \nn \\
&\times\MeijerG{M+2}{2}{3}{M+3}{k-n+1,\,0,\,1}{k+1+\nu_M,\,\ldots\,,\,k+1+\nu_1,\,0,\,0,\,k+1}{\gamma^{-1}}
\end{align}
Hence, the channel matrix does not depend on the ordering of the scattering objects as long as the signal passes through all scatterers.

\section{Conclusions and Outlook}
\label{sec:conclusion}

In this paper we have studied the correlations of the singular values of the product of $M$ rectangular complex matrices from independent Wishart ensembles. This generalises the classical result for the so-called Wishart--Laguerre unitary ensemble (or chiral unitary ensemble) at $M=1$, and is a direct extension of a recent result for the product of square matrices~\cite{AKW:2013}. We have seen that the problem of determining the statistical properties of the product of rectangular matrices can be equivalently formulated as a problem with the product of quadratic matrices and a modified, also called induced measure, see \cite{IK:2013} for a general derivation. The expense of this reformulation of the problem is the introduction of additional determinants in the partition function.

We have shown that the joint probability density function for the singular values can be expressed in terms of Meijer-$G$ functions. The approach which we have used relies on an integration formula for the Meijer-$G$ function as well as on the Harish-Chandra--Itzykson--Zuber integration formula. Due to the latter this method is limited to the complex case ($\beta=2$).
Furthermore, it has been shown, using a two-matrix model and the method of bi-orthogonal polynomials, that all correlation functions can be expressed as a determinantal point process containing Meijer-$G$ functions. From the explicit expressions we derived it follows that all correlation functions are independent of the ordering of the matrix dimensions.

The level density (or one-point correlation function) was discussed in detail. We used the spectral density to calculate all moments and derived its macroscopic limit. In particular, we analysed the location of the end points of the spectrum in the macroscopic limit for arbitrary $M$ and derived some narrow bounds for the location of these edges.

As an application we briefly discussed the ergodic mutual information, and how the singular values of products of random matrices are related to progressive scattering in MIMO communication channels.

The results presented in this work concern matrices of finite size, while previous results for the product of rectangular random matrices were only derived in the macroscopic large-$N_0$ limit. The explicit expressions for all correlation functions at finite size make it possible to also discuss microscopic properties, such as the local correlations in the bulk and at the edges. Due to known universality results for random matrices it is expected that such an analysis should reproduce the universal sine and Airy kernel in the bulk and at the soft edge(s), respectively, after an appropriate unfolding. Close to the origin the level statistics will crucially depend on whether or not the difference of the individual matrix dimensions to the smallest one, $\nu_m=N_m-N_0$, scales with $N_0$. If it does this will lead to a soft edge. Else it is expected, that the microscopic behaviour at the origin will be sensitive to $M$ and $\nu_m$. For a single matrix with $M=1$ (the Wishart--Laguerre ensemble), it is already known that this limit yields different Bessel universality classes labelled by $\nu_1$. 

Furthermore, the determinantal structure of the correlation functions make it possible to study the distribution of individual singular values, which is an intriguing problem in its own right.

It has been pointed out in~\cite{Zhang:2013}, that for the product of two square matrices, $M=2$ and $\nu_1=0$, the bi-orthogonal polynomials in question are special cases of multiple orthogonal polynomials associated with the modified Bessel function of the second kind. It is an intriguing task to see whether this approach can be extended to the more general case with $M\geq 2$ and rectangular matrices. Progress in this direction has already been made~\cite{KZ:2013}.

{\itshape Acknowledgments.}
We acknowledge partial support by SFB|TR12 ``Symmetries and Universality in Mesoscopic Systems''  of the German Science Foundation DFG (GA) and by
the International Graduate College IRTG 1132 ``Stochastic and Real World Models'' of the German Science Foundation DFG (J.R.I). Moreover we thank Arno Kuijlaars and Lun Zhang \cite{KZ:2013} as well as Eugene Strahov \cite{ES:2013} for sharing their private communications with us. 

\appendix

\section{Special Functions and some of their Identities}
\label{sec:meijer}

In this appendix we collect some definitions and identities for the generalised hypergeometric function and for the Meijer $G$-function, which are used in this paper.

The generalised hypergeometric function is defined by a power series in its region of convergence~\cite{GR:2000},
\begin{align}
\hypergeometric{p}{q}{a_1,\,\ldots\,,\,a_p}{b_1,\,\ldots\,,\,b_q}{z}\equiv \sum_{k=0}^\infty \frac{\prod_{i=1}^p(a_i)_k}{\prod_{i=1}^q(b_i)_k}\frac{z^k}{k!},
\label{meijer:hyper-def}
\end{align}
where the Pochhammer symbol is defined by $(a)_0=1$ and $(a)_n\equiv(a+n-1)(a)_{n-1}= a(a+1)\cdots(a+n-1)$ for $n\geq 1$. It is clear that the hypergeometric series~\eqref{meijer:hyper-def} terminates if any of the $a_i$'s is a negative integer. In particular, if $n$ is a positive integer then
\begin{equation}
\hypergeometric{p+1}{q}{-n,a_1,\,\ldots\,,\,a_p}{b_1,\,\ldots\,,\,b_q}{z}= \sum_{k=0}^n \frac{(-1)^kn!}{(n-k)!} \frac{\prod_{i=1}^p(a_i)_k}{\prod_{i=1}^q(b_i)_k}\frac{z^k}{k!},
\label{meijer:hyper-poly}
\end{equation}
which is a polynomial of degree $n$ or less.

The Meijer $G$-function can be considered as a generalisation of the generalised hypergeometric function. It is usually defined by a contour integral in the complex plane~\cite{GR:2000},
\begin{multline}
\MeijerG{m}{n}{p}{q}{a_1,\,\ldots\,,\,a_p}{b_1,\,\ldots\,,\,b_q}{z}\equiv \\
\frac{1}{2\pi\imath}\int_L du\,z^u\frac{\prod_{i=1}^m\Gamma[b_i-u]\prod_{i=1}^n\Gamma[1-a_i+u]}{\prod_{i=n+1}^p\Gamma[a_i-u]\prod_{i=m+1}^q\Gamma[1-b_i+u]}.
\label{meijer:meijer-def}
\end{multline}
The contour runs from $-\imath\infty$ to $+\imath\infty$ and is chosen such that it separates the poles stemming from $\Gamma[b_i-u]$ and the poles stemming from $\Gamma[1-a_i+u]$. Furthermore this contour can be considered as an inverse Mellin transform.
For an extensive discussion of the integration path $L$ and the requirements for convergence see~\cite{NIST:2010}.

It follows that the Mellin transform of a Meijer $G$-function is given by~\cite{GR:2000}
\begin{multline}
\int_0^\infty ds\,s^{u-1} \MeijerG{m}{n}{p}{q}{a_1,\,\ldots\,,\,a_p}{b_1,\,\ldots\,,\,b_q}{sz}=\\
z^{-u}\frac{\prod_{i=1}^m\Gamma[b_i+u]\prod_{i=1}^n\Gamma[1-a_i-u]}{\prod_{i=n+1}^p\Gamma[a_i+u]\prod_{i=m+1}^q\Gamma[1-b_i-u]},
\label{meijer:meijer-mellin}
\end{multline}
which is results from  the definition of the Meijer $G$-function~\eqref{meijer:meijer-def}. In combination with the definition of the gamma-function we have another identity
\begin{multline}
\int_0^\infty dt\, e^{-t}t^{b_0-1} \MeijerG{m}{n}{p}{q}{a_1,\,\ldots\,,\,a_p}{b_1,\,\ldots\,,\,b_q}{\frac{s}{t}}=\\
\MeijerG{m+1}{n}{p}{q+1}{a_1,\,\ldots\,,\,a_p}{b_0,\,\ldots\,,\,b_q}{s}.
\label{meijer:meijer-induc}
\end{multline}
Both of these integral identities are used throughout this paper. Another integral identity, which is used in section~\ref{sec:mimo}, allows us to integrate over the product of two Meijer $G$-functions~\cite{PBBM:1990},
\begin{multline}
\int_0^\infty ds\, \MeijerG{m}{n}{p}{q}{a_1,\,\ldots\,,\,a_p}{b_1,\,\ldots\,,\,b_q}{\eta s}\MeijerG{\mu}{\nu}{\sigma}{\tau}{c_1,\,\ldots\,,\,c_\sigma}{d_1,\,\ldots\,,\,d_\tau}{\omega s} = \\
\frac{1}{\omega}\MeijerG{m+\nu}{n+\mu}{p+\tau}{q+\sigma}{a_1,\,\ldots\,,\,a_n,\,-d_1,\,\ldots\,,\,-d_\tau,\,a_{n+1},\,\ldots\,,\,a_p}
{b_1,\,\ldots\,,\,b_m,\,-c_1,\,\ldots\,,\,-c_\sigma,\,b_{m+1},\,\ldots\,,\,b_q}{\frac{\eta}{\omega}}.
\label{meijer:meijer-product}
\end{multline}
The full set of restrictions on the indices for this integration formula can be found in~\cite{PBBM:1990}.

In addition to the integral identities given above, we need some other identities for the Meijer $G$-function. We employ several times that it is possible to absorb powers of the argument into the Meijer $G$-function, by making a shift in the arguments~\cite{GR:2000},
\begin{equation}
z^\rho\MeijerG{m}{n}{p}{q}{a_1,\,\ldots\,,\,a_p}{b_1,\,\ldots\,,\,b_q}{z}=\\
\MeijerG{m}{n}{p}{q}{a_1+\rho,\,\ldots\,,\,a_p+\rho}{b_1+\rho,\,\ldots\,,\,b_q+\rho}{z}.
\label{meijer:meijer-shift}
\end{equation}
For computing the function $\phi_n^M(s)$ in section~\ref{sec:2mm}, we need the differential identity~\cite{PBBM:1990}
\begin{multline}
z^n\frac{d^n}{dz^n}\MeijerG{m}{n}{p}{q}{a_1,\,\ldots\,,\,a_p}{b_1,\,\ldots\,,\,b_q}{\frac{1}{z}}= \\
(-1)^n\MeijerG{m}{n+1}{p+1}{q+1}{1-n,\,a_1,\,\ldots\,,\,a_p}{b_1,\,\ldots\,,\,b_q,\,1}{\frac{1}{z}}.
\label{meijer:meijer-diff}
\end{multline}
We also use that the generalised hypergeometric polynomial is related to the Meijer $G$-function by
\begin{multline}
\hypergeometric{1}{q}{-n}{b_1,\,\ldots\,,\,b_q}{z}=\\
n!\prod_{i=1}^q\Gamma[b_i]\,\MeijerG{1}{0}{1}{M+1}{n+1}{0,\,1-b_1,\,\ldots\,,\,1-b_q}{z},
\label{meijer:hyper-meijer}
\end{multline}
in order to write the polynomial $p_n^M(s)$ as a Meijer $G$-function in section~\ref{sec:2mm}.

As a last remark of this appendix, it should be mentioned that the Meijer $G$-function contains a vast number of elementary and special functions as special cases (e.g. see~\cite{Bateman:1953}). We mention that
\begin{equation}
\MeijerG{1}{0}{0}{1}{-}{b}{z}=z^be^{-z}
\quad\text{and}\quad
\MeijerG{1}{2}{2}{2}{1,\,1}{1,\,0}{z}={\rm ln}(1+z),
\label{meijer:meijer-func}
\end{equation}
which becomes useful in sections~\ref{sec:jpdf} and~\ref{sec:mimo}, respectively.


\raggedright

\end{document}